# Beyond Janus Atomic Ordering: High-Throughput First-Principles Search for Hidden MoSO Monolayer Structures

Zhijing Huang[1], Tingting Zeng[1], Lin Zhang[2], Zhibin Gao[1], Longyuzhi Xu[2] , Li Yang[1†], Shuming Zeng[2,†], Zonglin Gu[2,†]

[1]College of Physical Science and Technology, Guangxi Normal University, Guilin, Guangxi 541004, China

[2]College of Physical Science and Technology, Yangzhou University, Yangzhou, Jiangsu 225009, China

[†]Corresponding authors: guzonglin@yzu.edu.cn (Zonglin Gu), zengsm@yzu.edu.cn (Shuming Zeng), yangli@mailbox.gxnu.edu.cn (Li Yang).

## Abstract

Despite the growing interest in two-dimensional (2D) MoSO systems, existing studies have exclusively focused on conventional Janus structures. In this work, we perform high-throughput first-principles calculations to explore novel stable 2D MoSO monolayers. Combined with random sampling strategy, graph theory and group theory, we successfully screen out three novel non-Janus 2D MoSO monolayers from 1325 candidate structures, namely Reversed 2H-MoSO, Hybrid 2H-MoSO, and Hybrid 1T'-MoSO. Compared with Janus MoSO monolayers, the non-Janus MoSO counterparts possess lower binding energies, varying from −4.38 to −4.51 eV/atom. A systematic combination of dynamic, thermodynamic, and mechanical stability analyses corroborates their excellent structural robustness. *Ab* initio molecular dynamics (AIMD) simulations confirm their superior thermal resistance, with the structures remaining stable at temperatures beyond 2000 K. Interestingly, unlike the semiconducting Janus MoSO, the Hybrid 1T'-MoSO monolayer exhibits distinct metallic characteristics. Furthermore, we found that strain and curvature can enable controlled phase transitions of MoSO among semiconducting, semimetallic, and metallic phases. More importantly, the Hybrid 1T'-MoSO exhibits favorable HER activity with a Gibbs free energy of −0.002 eV, rendering it a promising candidate for hydrogen evolution catalysis. This work not only expands the family of 2D MoSO materials but also provides a reliable strategy for discovering stable functional 2D materials via high-throughput computation.

# I. INTRODUCTION

The precise design of interfacial structures and targeted property modulation for two-dimensional (2D) materials have long been central research subjects in surface and interface science[1-3]. Atomic-scale morphology fundamentally determines the intrinsic stability, electronic band characteristics and catalytic performance of layered 2D materials[4]. Minor alterations to stacking modes, in-plane atomic arrangements and surface atomic distributions can trigger significant variations in their physical and chemical properties[5,6]. Nevertheless, 2D materials possess an enormous structural space, bringing great challenges to experimental research. Conventional trial-and-error synthesis is time-consuming and expensive. A major challenge is that this approach struggles to effectively access many metastable and unreported atomic configurations with promising functional properties. Therefore, accurate theoretical prediction of interfacial structures is critical for achieving rapid experimental synthesis and targeted property modulation.

Transition metal dichalcogenides (TMDCs) represent a family of 2D layered materials with a typical sandwich structure, in which chalcogen layers tightly enclose a central layer of transition metal atoms[7-9]. As a typical TMDC, $MoS_2$ monolayer has been extensively explored due to its superior electronic, optical, and mechanical performances[10-15]. Experimentally, $MoO_3$ is commonly adopted as a molybdenum precursor for the synthesis of $MoS_2$ via chemical vapor deposition (CVD)[16-18]. However, incomplete sulfurization of $MoO_3$ during CVD growth inevitably yields molybdenum oxides ($MoO_{3-x}$) or molybdenum oxysulfides ($MoO_{3-x}S_y$) [19-22], implying the feasibility of synthesizing oxygen-incorporated MoSO monolayers. In 2020, Yagmurcukardes et al. theoretically demonstrated that Janus MoSO monolayers can be thermodynamically stable and exhibit prominent out-of-plane piezoelectricity[23]. Subsequent investigations further revealed the promising application potential of Janus MoSO in solar cells[24], optoelectronic devices[25], and spintronic systems[26]. In addition, Liu et al. recently developed a solution-based strategy to fabricate Janus 1T'-MoOSe by growing Se-Mo-O sandwiched shells on gold nanocores[27], providing an effective experimental approach for preparing oxygen-containing TMDCs monolayers. Despite these advances, most reported MoSO-related studies have focused exclusively on Janus asymmetric structures. Inspired by these experimental and theoretical advances, we propose that non-Janus MoSO monolayer structures may exist stably.

In this work, we systematically predict and analyze three stable non-Janus MoSO monolayer configurations via high-throughput first-principles calculations. The calculated results reveal their unique electronic properties and phase transitions modulated by strain and curvature. In particular, we find that the Hybrid 1T'-MoSO monolayer exhibits intrinsic metallic characteristics, which differs from the semiconducting nature of conventional Janus MoSO. To our knowledge, this is the first observation of a metallic 2D MoSO monolayer.

## II. COMPUTATIONAL DETAILS

The structure search for MoSO was performed using a stochastic approach based on group and graph theory, implemented with the $RG^2$ software[28]. Unreasonable structures were filtered out after the search. All calculations were carried out using first-principles density functional theory (DFT) within the Vienna Ab initio Simulation Package (VASP) [29]. The exchange–correlation interactions were treated using the generalized gradient approximation (GGA) with the Perdew–Burke–Ernzerhof (PBE) functional[30]. A plane-wave cutoff energy of 600 eV was adopted. For structural relaxation, an 11×6×1 k-point mesh was used to sample the Brillouin zone. Phonon dispersion calculations were conducted via density functional perturbation theory (DFPT) [31] using a 5×3×1 supercell. Elastic constants were derived using the stress-strain method to assess mechanical stability. *Ab* initio molecular dynamics (AIMD) simulations were performed on a 4×4×1 supercell to evaluate thermodynamic stability. For the HER calculations, the supercell size was set to 3×2×1. We used QVASP to expand the supercells of the three structures and roll them into nanotubes[32]. Band structure calculations were carried out along the *k*-path (Γ–X–S–Y–Γ) generated by VASPKIT[33]. The electron localization function (ELF) was visualized using the VESTA software[34]. A vacuum layer larger than 15 Å was applied along the *z*-direction to minimize interlayer interactions.s

## III. RESULTS AND DISCUSSION

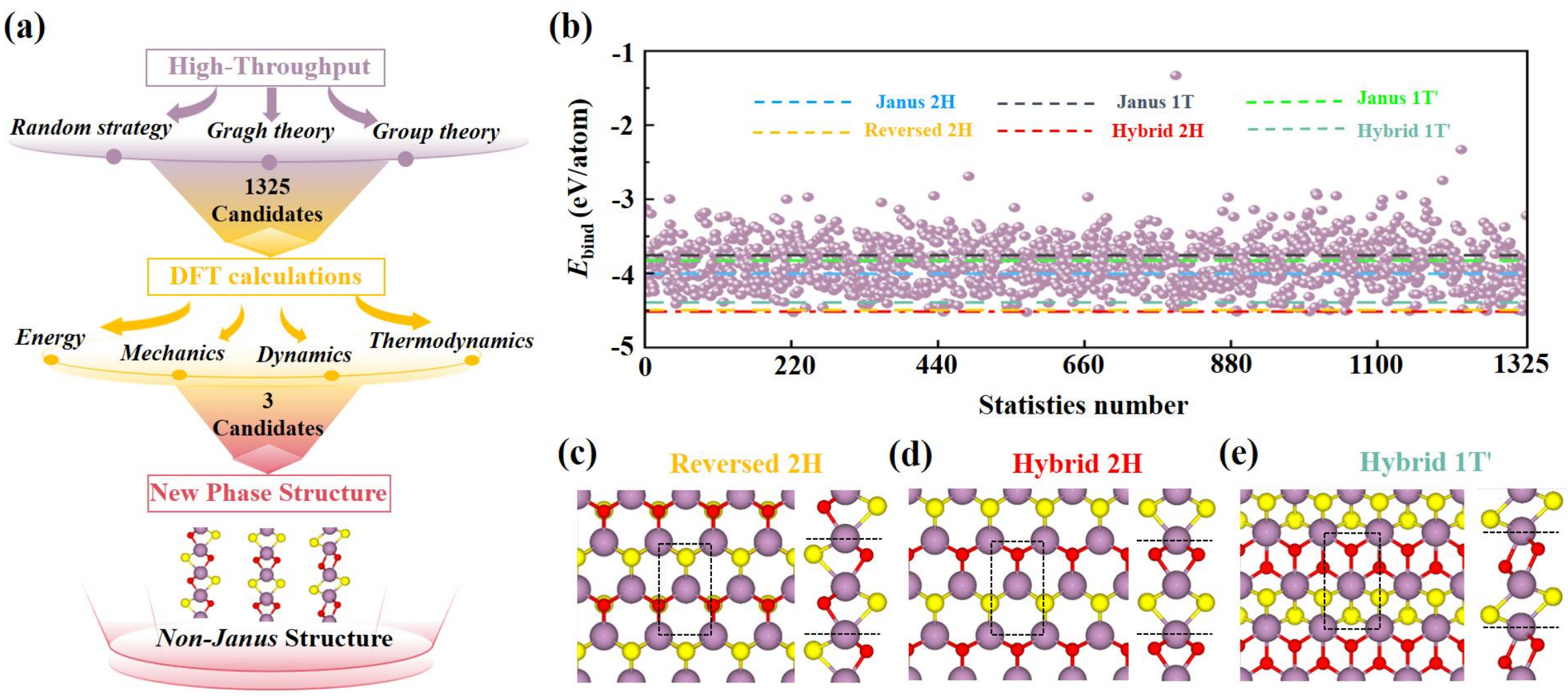

**FIG. 1.** (a) Schematic workflow for discovering stable non-Janus MoSO monolayers. (b) Binding energies ($E_{bind}$) of 1325 predicted MoSO monolayer structures, with dashed lines marking typical phases. Top (left) and side (right) views of non-Janus MoSO monolayers: (c) Reversed 2H-MoSO, (d) Hybrid 2H-MoSO, (e) Hybrid 1T'-MoSO. The purple, red and yellow spheres correspond to molybdenum (Mo), oxygen (O) and sulfur (S) atoms, respectively.

**Figure 1(a)** presents the workflow for predicting novel 2D MoSO monolayers via high-throughput screening and DFT calculations. First, a dataset of 1325 MoSO candidate structures was generated using a random strategy, graph theory, and group theory. Subsequently, DFT calculations were employed to analyze the stability of systems from four perspectives: energetic, mechanical, dynamic, and thermodynamic. Finally, structurally stable 2D MoSO monolayer configurations were identified from the dataset. For the energetic stability analysis, the binding energies ($E_{bind}$) were calculated using the formula: $E_{bind} = \left[E_{total} - \left(N \bullet E_{Mo}\right) - \left(N \bullet E_{S}\right) - \left(0.5 \bullet N \bullet E_{O2}\right)\right] / 3N$, where $E_{total}$ denotes the total energy of the MoSO monolayer, $E_{Mo}$ and $E_{S}$ are the energies of isolated Mo and S atoms, respectively, $E_{O2}$ is the energy of an oxygen molecule, $N$ represents the number of Mo, S, and O atoms (with equal counts of each atom type in the MoSO unit cell). A more negative value of $E_{bind}$ indicates higher energetic stability of the MoSO monolayer structure.

As shown in **Fig. 1(b)**, we present the statistical distribution of $E_{\mathrm{bind}}$ for all 1325 predicted MoSO monolayer candidates, covering a wide energy range from -4.51 to -1.2 eV/atom. Among these candidates, we identify not only the typical Janus phases (2H, 1T, and 1T') but also three stable non-Janus phases (Reversed 2H, Hybrid 2H, and Hybrid 1T'). Notably, in contrast to the Janus phase structures, these stable non-Janus configurations are concentrated in the low-energy region with $E_{\mathrm{bind}}$ of −4.38 eV/atom (Hybrid 1T'), −4.49 eV/atom (Reversed 2H), and −4.51 eV/atom (Hybrid 2H), respectively.

**Figure 1(c-e)** show the atomic structures of non-Janus MoSO monolayers in the Reversed 2H, Hybrid 2H, and Hybrid 1T' phases, respectively. For Reversed 2H-MoSO, S and O atoms are symmetrically distributed on both sides of the Mo atomic layer, forming a periodic alternating arrangement of O–Mo–S and S–Mo–O atomic configurations. For the Hybrid 2H/1T'-MoSO phase, the in-plane Mo atoms retain the characteristic hexagonal lattice. In contrast to the Janus and Reversed MoSO monolayers, both Hybrid MoSO structures can be regarded as atomic-level hybrids of $MoS_2$ and $MoO_2$ sublattices, as clearly seen in the side view of their unit cells. In Hybrid 2H-MoSO, the S and O atoms on both sides of the Mo atomic layer occupy the top sites, whereas those atoms preferentially reside at the hollow sites in Hybrid 1T'-MoSO. Notably, these three configurations are novel 2D MoSO monolayers with non-Janus structures, which have never been reported in previous studies. Their detailed structural parameters are listed in **Table SI**.

To evaluate the dynamic and thermodynamic stability of these novel 2D MoSO monolayers, we performed phonon spectrum calculations and AIMD simulations, respectively. As shown in **Fig. 2(a-c)**, the phonon dispersion curves of all three MoSO monolayers are free of imaginary frequencies across the entire Γ–X–S–Y–Γ high-symmetry path. All phonon spectra exhibit continuous positive branches throughout the Brillouin zone, with the highest optical modes at ~630 $cm^{-1}$ (Reversed/Hybrid 2H) and ~500 $cm^{-1}$ (Hybrid 1T'). The absence of soft modes confirms that all three structures are dynamically stable against small lattice perturbations. **Fig. S1** presents the Raman spectra and Raman-active vibrational characteristics of the three MoSO monolayers. Compared with Reversed 2H-MoSO and Hybrid 1T'-MoSO (**Fig. S1(a)** and **Fig. S1(e)**), Hybrid 2H-MoSO exhibits more prominent Raman peaks (Fig. S1(c)). The corresponding vibrational modes combine in-plane and out-of-plane vibrations (**Fig. S1(b)**, **Fig. S1(d)**, and **Fig. S1(f)**). Notably, the vibrational signals of O atoms are mainly concentrated in the high-frequency region, which originates from the much lower atomic mass of oxygen compared with molybdenum and sulfur atoms.

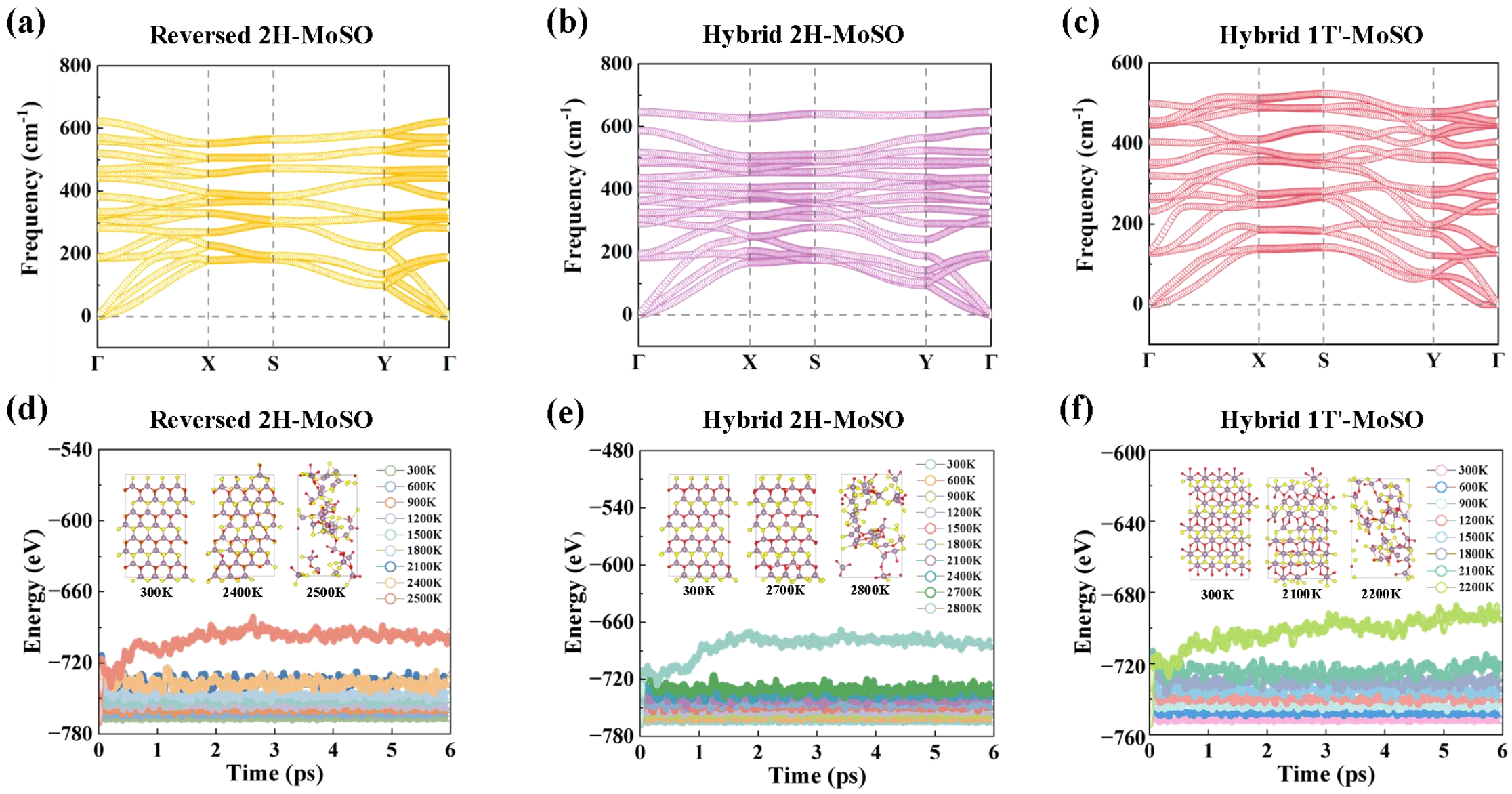

**FIG. 2.** (a-c) Phonon dispersion curves and (d-f) *ab* initio molecular dynamics (AIMD) simulation results of non-Janus MoSO monolayers. The inset snapshots show the top-view atomic structures of MoSO monolayers at 6 ps under different temperatures.

Thermodynamic stability was further assessed via AIMD simulations (**Fig. 2(d–f)**). For the Reversed 2H-MoSO monolayer, the total energy remains nearly constant with only small thermal fluctuations up to 2400 K, and the top-view atomic snapshot at 6 ps shows no significant structural degradation. At 2500 K, however, the energy exhibits a pronounced upward drift, accompanied by clear atomic disorder and lattice distortion in the corresponding snapshot, indicating the onset of thermal decomposition. The Hybrid 2H-MoSO monolayer maintains its structural integrity and stable total energy up to 2700 K, with the 6 ps snapshot showing a well-preserved crystalline lattice. At 2800 K, irreversible structural collapse occurs, as evidenced by the severely distorted atomic configuration and diverging total energy. The Hybrid 1T'-MoSO monolayer retains its ordered structure and stable energy profile up to 2100 K. Accordingly, the Hybrid 2H-MoSO exhibits the highest thermal tolerance at elevated temperatures.

**TABLE I.** Elastic constants of Reversed 2H-MoSO, Hybrid 2H-MoSO, Hybrid 1T'-MoSO, in GPa.

| MoSO | $C_{11}$ | $C_{12}$ | $C_{22}$ | $C_{66}$ | Stability |
|---|---|---|---|---|---|
| Reversed 2H | 85.49 | 26.62 | 85.09 | 30.57 | YES |
| Hybrid 2H | 86.94 | 24.46 | 80.86 | 28.57 | YES |
| Hybrid 1T' | 52.88 | 0.06 | 74.48 | 1.37 | YES |

Building on these stability assessments, we further evaluated the mechanical stability of the three novel MoSO monolayers by calculating their elastic constants, as summarized in **Table I**. For the MoSO monolayers, the Born stability criteria impose the following necessary conditions for mechanical stability[35]: $C_{11}C_{22} > C_{12}C_{12}$ and $C_{66} > 0$. As shown in **Table I**, all calculated elastic constants for the three MoSO phases satisfy these criteria. Consequently, all three MoSO monolayers are mechanically stable, consistent with their dynamical and thermodynamic stability confirmed by phonon spectrum calculations and AIMD simulations. We further investigated the mechanical properties of these MoSO monolayer structures by calculating the strain energy ($\Delta E$), Young's modulus ( $Y(\theta)$ ), and Poisson's ratio ( $\upsilon(\theta)$ ), as illustrated in **Fig. S2**. The $\Delta E$ is defined as: $\Delta E = E_{\varepsilon_x=i,\, \varepsilon_y=j} - E_{\varepsilon_x=0,\, \varepsilon_y=0}$, where $E_{\varepsilon_x=i,\, \varepsilon_y=j}$ is the total energy of the strained system under biaxial strain conditions (with $i$ and $j$ ranging from -10 to 10), and $E_{\varepsilon_x=0,\, \varepsilon_y=0}$ corresponds to the total energy of the unstrained pristine structure. The $Y(\theta)$ and $\upsilon(\theta)$ were subsequently derived according to the following equations:

$$Y(\theta)=\frac{C_{11}C_{22}-C^2{}_{12}}{C_{22}(\cos\theta)^4+A(\cos\theta)^2(\sin\theta)^2+C_{11}(\sin\theta)^4}$$

$$\upsilon(\theta)=\frac{C_{12}(\cos\theta)^4-B(\cos\theta)^2(\sin\theta)^2+C_{12}(\sin\theta)^4}{C_{22}(\cos\theta)^4+A(\cos\theta)^2(\sin\theta)^2+C_{11}(\sin\theta)^4}$$

where $A=\frac{C_{11}C_{22}-C^2{}_{12}}{C_{66}}-2C_{12}$, $B=C_{11}+C_{22}-\frac{C_{11}C_{22}-C^2{}_{12}}{C_{66}}$, and $\theta$ is the angle to the axes of 100. The elliptical distribution of $\Delta E$ and noncircular profiles of the $Y(\theta)$ and $\upsilon(\theta)$ demonstrate the anisotropic mechanical characteristics of the three MoSO monolayers. In terms of mechanical stiffness and flexibility, the three MoSO monolayers exhibit comparable mechanical performance. Notably, Hybrid 2H-MoSO delivers the maximum $Y(\theta)$ of 490 $N{\cdot}m^{-1}$ and a $\upsilon(\theta)$ of 0.995 among the three MoSO structures.

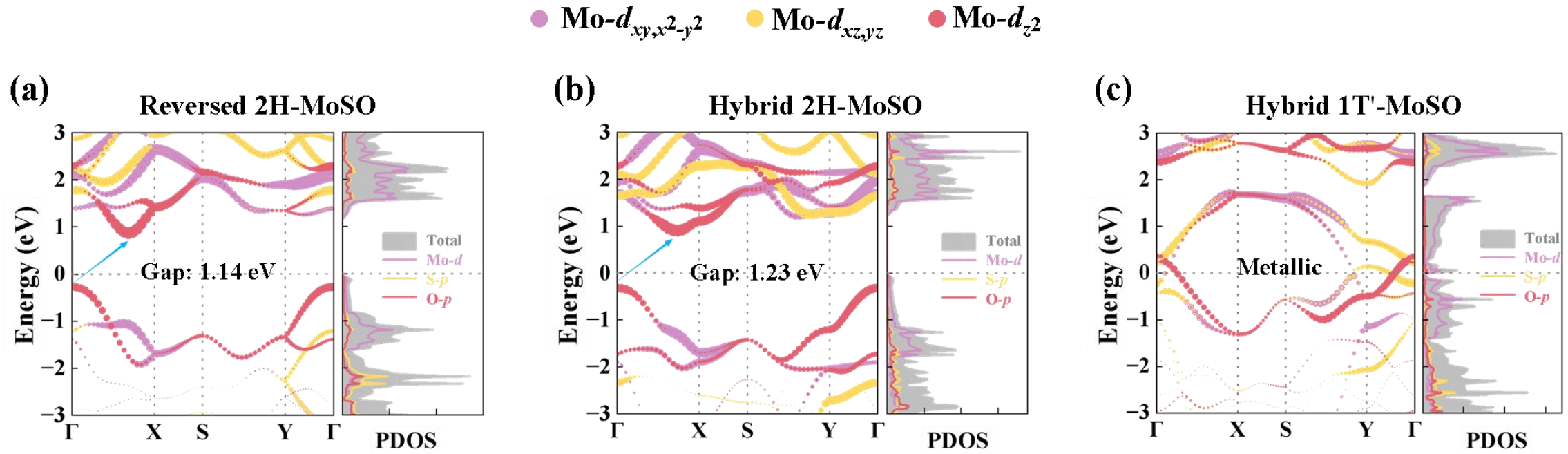

**FIG. 3.** The orbital-resolved band structures and projected density of states (PDOS) for (a) Reversed 2H-MoSO, (b) Hybrid 2H-MoSO, (c) Hybrid 1T'-MoSO. The purple, yellow, and red lines highlight the $d_{xy,x^2-y^2}$, $d_{xz,yz}$, and $d_{z^2}$ orbitals of the Mo atom, respectively. The Fermi level is set to zero.

To explore the electronic structure of these novel MoSO monolayers, we calculated their orbital-resolved band structures and projected density of states (PDOS), as shown in **Fig. 3**. The Fermi level is set to zero for all cases. As shown in **Fig. 3(a)** and **Fig. 3(b)**, the Reversed 2H-MoSO and Hybrid 2H-MoSO are semiconductors with indirect band gaps of 1.14 eV and 1.23 eV, consistent with the semiconducting behavior of known Janus MoSO phases (typically with band gaps of ~1.62 eV). Notably, the Hybrid 1T'-MoSO monolayer displays fundamentally different electronic behavior, as shown in **Fig. 3(c)**. In contrast to the semiconducting 2H phases, its band structure exhibits multiple bands crossing the Fermi level, which is a clear signature of metallic character. For both Reversed 2H-MoSO and Hybrid 2H-MoSO, the valence band maximum (VBM) and conduction band minimum (CBM) are primarily contributed by Mo-$d_{z^2}$ orbitals. For the metallic Hybrid 1T'-MoSO, two distinct Dirac points emerge near the Y point. One is located on the left side of the Y point, where the electronic states near the Fermi level are predominantly contributed by the Mo-$d_{xy,x^2-y^2}$ and Mo-$d_{xz,yz}$ orbitals. The other lies on the right side of the Y point, and the electronic states are mainly derived from the Mo-$d_{z^2}$ and Mo-$d_{xz,yz}$ orbitals. To further clarify the orbital origins of these distinct electronic behaviors, we turn to the PDOS analysis. From the PDOS spectra, we observe that the electronic states near the band edges (for the semiconducting phases) and at the Fermi level (for the metallic phase) are overwhelmingly dominated by Mo-*d* orbitals, with only minor contributions from S-*p* and O-*p* orbitals. We further carried out systematic analyses on the electron localization function (ELF), Bader charge, crystal orbital Hamilton population (COHP), and work function of the three MoSO monolayers. As shown in **Fig. S3**, the Mo–O and Mo–S bonds within these monolayers are dominated by covalent bonding features. Bader charge results demonstrate that charge transfers from Mo atoms to neighboring O and S atoms (see **Fig. S4**). COHP calculations suggest that Reversed 2H-MoSO has the strongest Mo–O bond strength, whereas Hybrid 2H-MoSO owns the most robust Mo–S bonds (see **Fig. S5**). **Fig. S6** compares the work functions of the three MoSO monolayers. Relative to Reversed 2H-MoSO and Hybrid 2H-MoSO, Hybrid 1T'-MoSO shows a higher work function of 5.7 eV.

Building on the electronic structure analysis of the three MoSO phases, we next investigate the strain-induced tunability of their band structure and phase transitions, as illustrated in **Fig. 4**. Panels (a), (d), and (g) present surface plots of the bandgap evolution as a function of in-plane biaxial strain for Reversed 2H-, Hybrid 2H-, and Hybrid 1T'-MoSO, respectively. These contour maps reveal that the bandgap is highly sensitive to strain magnitude and direction, with distinct responses across the three phases. As shown in **Fig. 4(a),** the indirect bandgap of Reversed 2H-MoSO decreases progressively under increasing tensile strain, eventually driving the system into a metallic state. This transition is explicitly demonstrated by the band structures in **Fig. 4(b)** and **Fig. 4(c)**. At the strain state of ($\varepsilon_x$ = 4 %, $\varepsilon_y$ = 6%), Reversed 2H-MoSO remains an indirect-bandgap semiconductor with a reduced gap of 0.207 eV. In contrast, at ($\varepsilon_x$ = 6 %, $\varepsilon_y$ = 6 %), the bands cross the Fermi level, confirming the transition to a metallic phase. Similarly, as shown in **Fig. 4(d)**, the bandgap of Hybrid 2H-MoSO gradually narrows with increasing tensile strain, first evolving into a semimetallic state and then fully transitioning to a metallic phase.

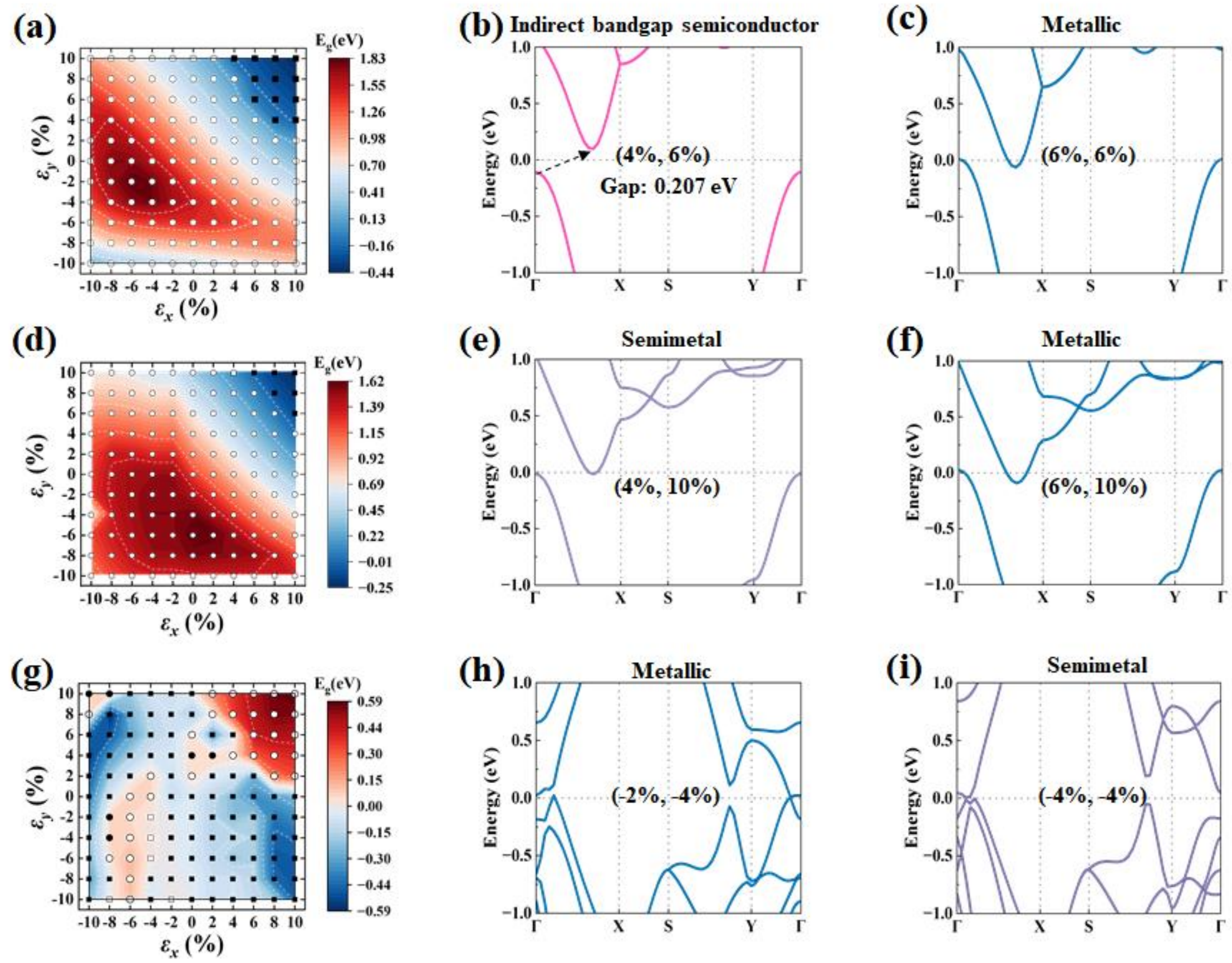

**FIG. 4.** The surface plot of in-plane biaxial strain ($\varepsilon_x$, $\varepsilon_y$) versus bandgap evolution in (a) Reversed 2H-MoSO, (d) Hybrid 2H-MoSO and (g) Hybrid 1T'-MoSO. Strain-modulated electronic band structures of (b-c) Reversed 2H-MoSO at $\varepsilon_x$ = 4% , 6% and $\varepsilon_y$ = 6%, (e-f) Hybrid 2H-MoSO at $\varepsilon_x$ = 4%, 6% and $\varepsilon_y$ = 10%, (h-i) Hybrid 1T'-MoSO at $\varepsilon_x$ = -2%, -4% and $\varepsilon_y$ = -4%. Notions: white/black circles represent indirect/direct bandgap semiconductors; white/black squares indicate semimetals/metallic phases.

The corresponding electronic structures are presented in **Fig. 4(e)**and **Fig. 4(f)**. At ($\varepsilon_x$ = 4 %, $\varepsilon_y$ = 10 %), the conduction and valence bands touch at the Fermi level, forming a semimetallic state. At ($\varepsilon_x$ = 6 %, $\varepsilon_y$ = 10 %), multiple bands cross the Fermi level, verifying the complete transition to a metallic phase. In contrast, the metallic Hybrid 1T'-MoSO shows a distinct response to compressive strain, as shown in **Fig. 4(g)**. The contour plot reveals that compressive strain significantly modifies its electronic structure, inducing semimetallic features in specific strain regimes. This behavior is directly visualized in the band structures in **Fig. 4(h)** and **Fig. 4(i)**. At ($\varepsilon_x$ = -2 %, $\varepsilon_y$ = -4 %), the Hybrid 1T'-MoSO retains its metallic character with multiple band crossings at the Fermi level. A more compressive strain of ($\varepsilon_x$ = -4 %, $\varepsilon_y$ = -4 %), the bands approach the Fermi level without fully crossing it, leading to a semimetallic state with a small overlap between the conduction and valence bands. These results demonstrate that in-plane biaxial strain serves as an effective tool to continuously tune the electronic properties of these MoSO monolayers, enabling controlled transitions between semiconducting, semimetallic, and metallic phases.

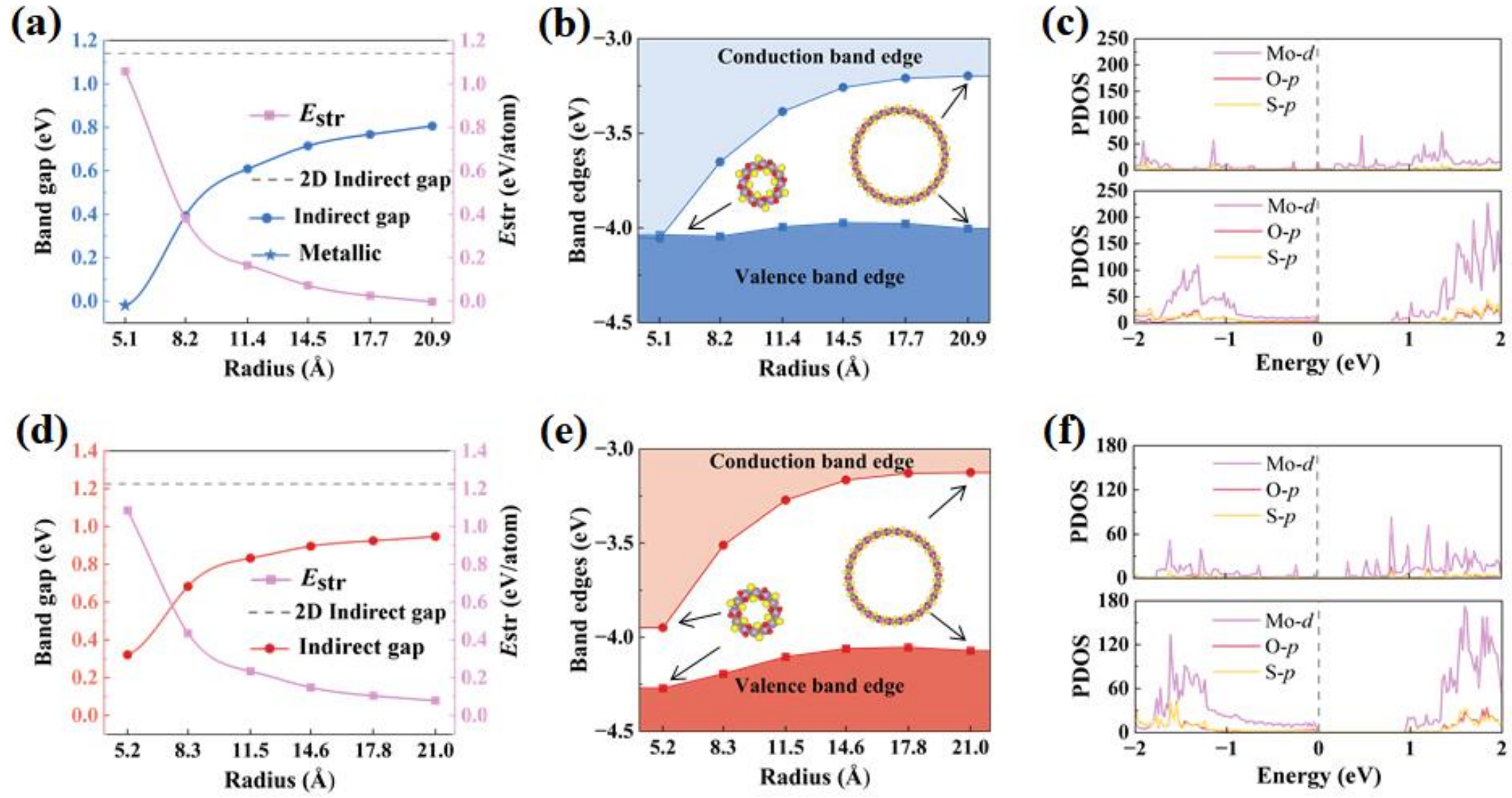

**FIG. 5.** (a,d) Strain energy ($E_{str}$) and band gaps versus the nanotube radius for Reversed/Hybrid 2H-MoSO. The indirect band gap of 2D MoSO is indicated by the grey dashed line. (b,e) Variation of valence and conduction band edge positions with nanotube radius. Insets illustrate nanotube geometries at selected diameters, with the corresponding projected density of states (PDOS) provided in (c,f).

**Fig. 5(a)** and **Fig. 5(d)** illustrate the variations of strain energy and band gap with nanotube radius for two types of nanotube structures (see the structures in **Fig. S8** and **Fig. S9**). The strain energy is calculated using the formula $E_{str} = (E_{NT}/N_{NT}-E_M/N_M)$, where $E_{NT}$ and $E_M$ are the total energies of the nanotube and the 2D monolayer, respectively, and $N_{NT}$ and $N_M$ are the corresponding numbers of unit cells. A lower strain energy indicates a more stable structure. From **Fig. 5(a)** and **Fig. 5(d)**, we observe that for both the Reversed 2H-MoSO and Hybrid 2H-MoSO structures, the strain energy decreases with increasing radius, while the band gap increases (see the band structures in **Fig. S9** and **Fig. S10**). It is worth noting that Hybrid 1T'-MoSO nanotubes retain metallic characteristics as the tube diameter increases (**Fig. S11** and **Fig. S12**). **Fig. 5(b)** and **Fig. 5(e)** further show the evolution of the conduction band and valence band with radius. It is found that the valence band edges of both structures exhibit little change, whereas the conduction band edges shift significantly with increasing radius, indicating that the increase in band gap is mainly contributed by the conduction band. Notably, the Reversed 2H-MoSO structure undergoes a phase transition: at a radius of 5.1 Å, it is metallic, and as the radius increases, it transforms into a semiconductor. **Fig. 5(c)** and **Fig. 5(f)** display the projected density of states (PDOS) for selected radii of the Reversed 2H-MoSO and Hybrid 2H-MoSO structures, respectively. The PDOS further confirms that the band gap variation is primarily governed by the conduction band, with the Mo-*d* orbitals providing the main contribution, followed by the O-*p* and S-*p* orbitals.

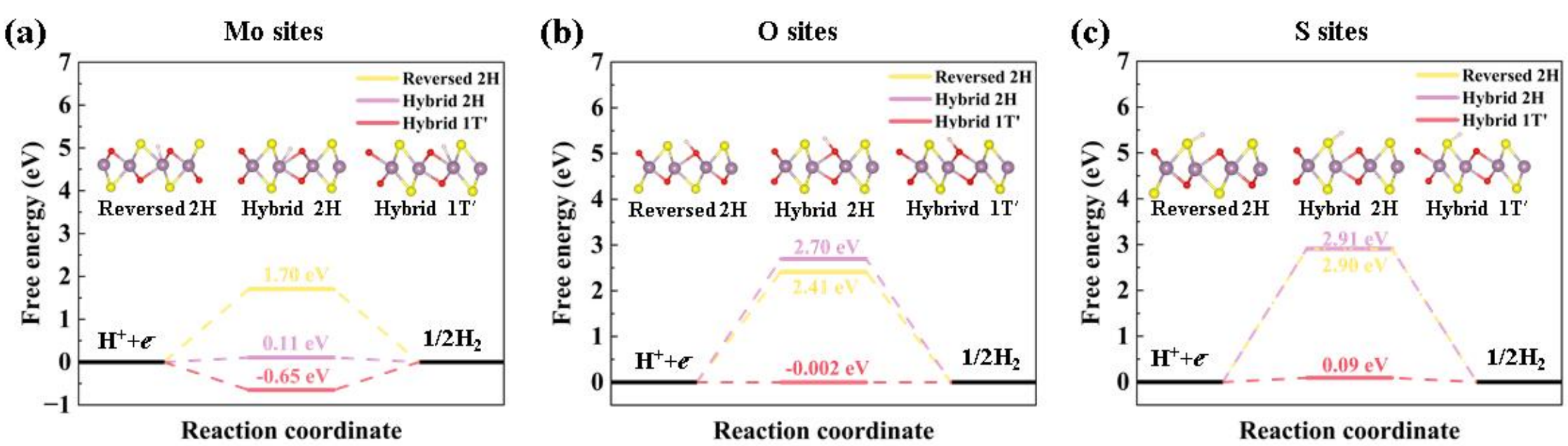

**FIG. 6.** Free energy of the hydrogen evolution reaction (HER) for Reversed 2H-MoSO, Hybrid 2H-MoSO, and Hybrid 1T'-MoSO with H adsorption on (a) Mo sites, (b) O sites, and (c) S sites. Insets show the atomic configurations of H adsorption at different sites.

The hydrogen evolution reaction (HER) activity of the three MoSO monolayers is evaluated via the Gibbs free energy of hydrogen adsorption ($\Delta G_{H^*}$), as shown in **Fig. 6**. Panels (a)-(c) display the corresponding HER free-energy diagrams for H adsorption at Mo, O, and S sites, with the adsorption geometries provided in the insets. At the Mo site, the calculated $\Delta G_{H^*}$ values are 1.70 eV, 0.11 eV, and -0.65 eV for Reversed 2H-, Hybrid 2H-, and Hybrid 1T'-MoSO, respectively. Notably, Hybrid 1T'-MoSO exhibits a near-optimal $\Delta G_{H^*}$ of -0.002 eV at the O site, satisfying the thermoneutral condition for efficient HER. At the S site, Reversed 2H- and Hybrid 2H-MoSO show large positive $\Delta G_{H^*}$ values of 2.90 eV and 2.91 eV, respectively, while Hybrid 1T'-MoSO yields a small value of 0.09 eV. These results demonstrate that Hybrid 1T'-MoSO exhibits favorable HER activity at multiple adsorption sites, making it a promising candidate for hydrogen evolution catalysis.

## IV. CONCLUSIONS

In summary, a high-throughput first-principles search identified three stable non-Janus MoSO monolayers, namely Reversed 2H-MoSO, Hybrid 2H-MoSO, and Hybrid 1T'-MoSO, from 1325 candidate structures. All three phases exhibit energetic, dynamic, mechanical, and thermal stability. Reversed 2H-MoSO and Hybrid 2H-MoSO are indirect-gap semiconductors, whereas Hybrid 1T'-MoSO is metallic. Strain and curvature enable transitions among semiconducting, semimetallic, and metallic states. Moreover, Hybrid 1T'-MoSO exhibits nearly thermoneutral hydrogen adsorption with $\Delta G_{H^*} = -0.002$ eV, indicating promising HER activity. These findings extend the structural space of MoSO beyond conventional Janus configurations and demonstrate atomic arrangement as an effective means of tuning the electronic and catalytic properties of two-dimensional materials.

**Acknowledgements**

This research was supported by the School-level Cultivation Project of the National Natural Science Foundation of China in 2025, No. RZ2500003755, the National Natural Science Foundation of China under Grants No. 12204402, Talent Special Project, No. DC2400003207, the Natural Science Foundation of Guangxi Province (No. 2022GXNSFAA035487), and the Hefei Advanced Computing Center.